# Implementation of Advanced Wind Turbine Controllers for Scaled Turbine Testing in a Wind Tunnel

M Sinner[a], V Petrović[b], and L Y Pao[a]

[a]University of Colorado Boulder
[b]ForWind, University of Oldenburg

E-mail: michael.sinner@colorado.edu

**Abstract**

Based on a series of two experimental campaigns testing advanced controllers on a scaled wind turbine operating in a wind tunnel, this contribution describes the overall experimental method, challenges faced, lessons learned, and opportunities for future work. The two campaigns, run in Fall 2018 and Fall 2019, tested unconstrained and constrained optimal blade pitch controllers, respectively, using preview disturbance measurements of the oncoming wind. Specifically, the first study considered an extension to the linear-quadratic regulator to include feedforward action, while the second deployed model predictive control to incorporate actuator constraints into the optimal control problem. The results of the campaigns have already been published in technical conference and journal papers on control systems; however, detail on how the controllers were implemented was not included in those works. We aim to fill that gap with this contribution, which is targeted at the wind energy community. We describe several aspects of the experimental setup, in particular providing details of the software and hardware used for the controller; share insight on several aspects of the procedure that were difficult and how we overcame those challenges; and summarize the key differences between simulation-based studies and physical testing. By doing so, we hope to share what we learned during our experimental campaigns and provide a point of reference for others looking to carry out experiments on scaled wind turbines operating in wind tunnel facilities.

*Keywords*: Scaled wind turbine model, hardware in the loop, active grid, feedforward control

## 1 Introduction

Over the past 20 years or so, as the deployment of wind energy across the globe has exploded [1] and the hunt for a lower and lower cost of renewable energy continues, a large body of literature exploring new methods for wind turbine control has developed. The majority of academic research focuses on two main actuators for turbine control: the electrical load provided by the turbine generator (referred to as "generator torque control") and the pitch angle of the turbine blades ("blade pitch control"). Generally speaking, generator torque control is used to maximize the power generated by the turbine when wind speeds are low, while at higher wind speeds, blade pitch control is used to regulate the turbine to its rated speed while the generator torque is fixed at its rated torque, thus producing the rated power of the turbine [2].

A full review of the literature on wind turbine control designs is out of the scope of this paper; we will briefly mention a few broad classes and direct the reader to Laks et al.'s survey on advanced control [3], Scholbrock et al.'s survey on lidar-enabled control [4], and Lio et al.'s survey on model predictive control [5] for further reading. The literature may be broadly divided into research into advanced control algorithms that use existing sensors and actuators, including optimal multi-input multi-output controllers, among others; development of new actuation techniques, of which individual pitch control has received particular interest for its ability to reduce structural loads on the turbine while maintaining a high power output [6]; and control approaches that make use of extra sensors, most notably lidar-enabled feedforward control [7, 4], which uses measurements of the wind upstream of the turbine to 'pre-actuate' and reduce the impact of disturbances; as well as combinations of the above. Over the years, many promising results have been reported from simulation-based studies indicating that advanced controls can increase power production and decrease turbine structural loading.

Despite numerous studies with positive results, it appears that industry has been sluggish to adopt new controllers. To the best of our understanding, most turbine manufacturers still implement a series of single-input, single-output control loops to achieve the desired behavior with the blade pitch and generator torque controllers designed largely independently; for an example of an industry-standard controller, see Jonkman et al.'s NREL 5MW reference turbine controller, which was designed to reflect industry practices [8, Section 7]. A likely explanation for this is the lack of physical implementations of advanced wind turbine controllers—the vast majority of results have been reported using computer simulations. Although several excellent high- and medium-fidelity turbine modelling packages exist for the express purposes of simulating the nonlinear wind turbine response (FAST [9], for example), they of course cannot capture all of the relevant physics and as such, may not be as convincing as a physical test to turbine manufacturers. On the other hand, testing novel controllers on utility-scale turbines can be a tall order for academics and researchers, due to the high cost and potential risk of utility-scale experimental campaigns. Some studies have been reported both on research turbines [6, 10] and from manufacturers [11]; however, due to the aforementioned barriers and also the time it can take to implement advanced controls for research purposes, many of the controllers reported in simulation-based work have yet to be tested on physical turbines.

While deployment on utility-scale wind turbines remains the gold standard for demonstrating controllers, testing on scaled wind turbine models presents an opportunity to validate control algorithms on physical turbines without the high costs and risks associated with experimenting on utility-scale turbines. Testing controllers on scaled wind turbine models bypass many of the barriers to utility-scale testing. The models are (relatively) low-cost (although the cost of the testing facility may be high) and present a lower safety risk, while generally being quicker to deploy and test controllers on. Moreover, they can be tested in controlled environments, meaning that various different control designs can be evaluated using the same set of inflow conditions, which is generally impossible to achieve using a utility-scale turbine operating in the atmospheric boundary layer. Scaled-model testing is not without its disadvantages: it is not generally possible to scale down all of the relevant structural and aerodynamic properties of a utility-scale wind turbine, meaning that the scaled model is still not a completely accurate reflection of all of the relevant physics. Obtaining representative Reynolds numbers is particularly difficult to achieve [12]. Time-scaling also presents both challenges and opportunities: while the time speed-up means that testing that would require hours on a utility-scale turbine can be completed in minutes on a scaled model, it also means that control algorithms should ideally be sped up, which is not always possible (in our tests [13, 14], we used a controller sample rate comparable to that of a utility-scale turbine, which made the controller relatively slow for the scaled model). As a result, scaled-model tests cannot completely replace utility-scale testing, but can be used as an important proving ground for physically-implemented controllers.

Such scaled wind turbine tests have been successfully carried out at various wind tunnel facilities around the world. Amongst others, the Delft University of Technology and the Politecnico di Milano operate tunnels with experimental setups for both individual turbine control [15] and wind farm control [16]; and the University of Texas have recently commissioned a wind tunnel that can be used for scaled wind turbine testing, among other applications. Further, recent interest in floating offshore wind turbines has driven the development of combined wind tunnel/wave tank testbeds such as that at the University of Maine. However, while there are a few papers that briefly describe the software used [17, 18], the literature on scaled-model testing focuses on the scientific results and does not tend to provide details on the controller implementation process.

The experiments upon which this paper is based were carried out at ForWind – Center for Wind Energy Research in the Institute of Physics at the University of Oldenburg in Oldenburg, Germany using the 1.8 m rotor diameter Model Wind Turbine Oldenburg (MoWiTO). The wind tunnel and experimental setup will be described in Section 2.3—for further details of the controller testbed, please refer to Petrović et al. [19]. We conducted two experimental campaigns, the first in the fall of 2018 [13] and the second in the fall of 2019 [14]. The first considered a preview-enabled version of the linear-quadratic regulator, and the second extended the corresponding optimal control problem to include blade pitch actuator constraints, resulting in a linear model predictive controller. The technical details and results from these studies can be found in our previous publications [13, 14]; the purpose of the present paper is to describe the steps taken to implement the controllers on the scaled-model turbine and our takeaways from the experiments. To that end, we describe the test setup, software development, and validation steps in Section 2 and present our conclusions and recommendations to others planning scaled-model testing of advanced turbine controllers in Section 3.

| Simulation-based studies | Physical testing on the MoWiTO |
|---|---|
| **1.** Generate a simplified (often linear) mathematical model of the wind turbine to use for controller design purposes using either a physics-based derivation or numerical identification approaches. | **1.** Identify a linear mathematical model of the scaled-model wind turbine using a system identification applied to input-output data from a higher-order FAST model of the turbine. |
| **2.** Construct a controller based on the simplified mathematical model according to control system design principles (see, for example, Franklin et al. [20] for a thorough text on the fundamentals of control system design and Laks et al. [3] and Pao and Johnson [2] for reviews of advanced controllers in wind turbine applications). | **2.** Construct a linear quadratic regulator-type controller (with a feedforward extension) based on the simplified mathematical model. |
| **3.** Test the controller in closed loop using a higher-fidelity (often nonlinear) simulation model of the turbine as the 'true' system to validate performance. Tune the controller based on the simulated response and repeat the simulations as needed. | **3.** Test the controller in a closed-loop simulation using the FAST model to represent the true behavior of the turbine. Retune the controller as needed. |
|  | **4.** Implement the controller on the final control software and hardware, and verify behavior of controller implementation. |
| **5.** Test the controller in closed loop using the higher-fidelity simulation truth model and various randomly-generated inflow with various random seeds to validate the controller fully. Often, a reference controller representing industry standards is also simulated as a point of comparison. | **5.** Test the controller in closed loop with the scaled-model turbine using various repeated wind inflows. Test various controller configurations back-to-back to ensure consistency in the testing scenario. |

Table 1: Summary comparison of the general approach used in simulation-based studies and the physical testing we conducted.

## 2 Experimental approach

As mentioned previously, the stated purpose of the experimental campaigns that we conducted was to demonstrate the effectiveness of advanced feedforward control techniques [13, 14] and, in particular, model predictive control [14] on a physical wind turbine in order to validate simulation-based results reported in a large body of literature. The experimental approach that we used differs from the approach used in many simulation-based studies; these differences are recorded in Table 1.

The approach we took for physical testing depended on the fact that we had a FAST model of the MoWiTO available for simulation. If no higher-order truth model is available for simulation, step 3 may need to be replaced with initial testing on the true physical system. Further, in that case, a data-driven system identification procedure could be used to generate a mathematical model for step 1. Often, for both simulation-based and physical studies, steps 2 and 3 may be repeated several times, with the controller being retuned in step 2 in order to improve its performance. In some cases, step 1 might also be repeated in the design loop if the mathematical model itself is deemed to be a source of poor performance.

We will now describe the key steps of the experimental approach used for physical testing in more detail. In reporting on our procedure, we hope to provide a reference to other researchers looking to implement wind turbine controllers on scaled-model testbeds.

## 2.1 Software development

We used Matlab [21] for controller development. The main reason for this is that there is a FAST interface for Simulink, making testing controllers in closed-loop simulations particularly easy using Matlab/Simulink. On top of that, qpOASES [22], which we used to solve the model predictive control problem online [14], has a convenient Matlab interface. If the controller can be represented by a relatively simple differential (or difference) equation (such as a linear state-feedback gain [13]), it may be easiest to implement this directly in Simulink using elementary blocks. On the other hand, if the controller is algorithmic (as is the case with model predictive control [14]), text-based code may be needed. In our Simulink implementation, we used a Level-2 Matlab S-Function to pass the controller inputs to a class object containing the controller code before returning the controller output (control input to the plant). This approach has the added benefit that several controllers can be simulated with no change to the Simulink model, simply by creating instances of different controller classes with similar method names. The process of simulating the controller on a higher-fidelity truth model (in our case, a FAST model) may be thought of as 'software in the loop' testing.

Once the controller had been designed and tuned (steps 2 and 3 of the testing procedure in Table 1), we then translated our Matlab/Simulink code into LabVIEW [23] for real-time use on the physical MoWiTO. We constructed our controller as a LabVIEW virtual instrument (VI) that took as inputs the feedback and feedforward signals at time index $k$ and produced as an output the control action at time index $k$. This VI contained both the elementary control law but also a number of simple subVIs that handled controller output (plant control input) saturation, internal state integration, and fault handling in the case of the model predictive controller.

Relatively simple controllers [e.g. 13] can be implemented directly using LabVIEW blocks. However, algorithmic controllers such as model predictive controllers may need to be implemented using text-based code. To achieve this, we translated our model predictive controller code into C++; this essentially resulted in a MoWiTO-specific C++ wrapper for qpOASES, which is also written in C++. We compiled this code into a shared library that could be loaded and initialized prior to run-time, and interfaced with the LabVIEW controller using the LabVIEW-provided "Call Library Function" node. We used the node in the "Run in any thread" configuration—see Section 2.2.

## 2.2 Controller validation

Step 4 in Table 1 is perhaps the most important difference between simulation-based and physically implemented controller tests. In order to avoid damage to the physical system, it is important to validate the controller implementation. We did this in two steps: the first to validate the control software and the second to validate the control hardware.

To ensure that our translation of the Simulink/Matlab simulation controller to the LabVIEW/C++ real-time control software was accurate, we looked at the input/output behavior of the real-time control software in open-loop using controller input data from the Simulink model simulations. To do so, we developed a wrapper for our controller VI that read in controller input data generated during Simulink simulations, passed the data to the controller VI, and recorded the outputs. We could then compare the LabVIEW-based controller outputs to the Simulink-simulated controller outputs to verify that the controller software was behaving correctly and producing the expected input/output behavior.

The validation just mentioned looked only at the input/output behavior of the software in open loop, and importantly, did not include any real-time control hardware. A crucial step for validation is to check that the control software still runs as expected when implemented on the control hardware: in our case, a National Instruments CompactRIO (cRIO, see Section 2.3). Implementing the LabVIEW code onto the control hardware was straightforward, since both LabVIEW and the cRIO are National Instruments products; however, compiling the C++ code for the cRIO target (which runs on a version of Linux) was not entirely straightforward. After much trial and error, we were able to correctly compile the code using "C & C++ Development Tools for NI Linux Real-Time 2014, Eclipse Edition", a version of the Eclipse integrated development environment tailored by National Instruments. We then checked the implementation using a hardware-in-the-loop (HIL) test environment developed by Syed Muzaher Hussain Shah at the University of Oldenburg. This setup allowed us to run the simulation FAST model in closed loop with the controller on the cRIO using a specially-developed shared library to handle communications between the FAST executable and cRIO-based controller.

Th HIL testing step proved extremely useful. We found that the solve time for qpOASES (the optimization package used to solve our model predictive control problem) was somewhat slower on the cRIO than it had been on a desktop computer, and we were able to adjust the prediction horizon length accordingly to make up for this difference [14]. Moreover, we identified during HIL testing that we should be using the "Run in any thread" (rather than the default "Run in UI thread") configuration for the Call Library Function node in our controller, which proved vital to achieving acceptable controller speed and behavior.

### 2.3 Physical testing

The MoWiTO (the scaled-model turbine used in our experiments) is designed as an aerodynamically scaled model of the NREL 5MW reference turbine [8] (although the blades and tower are stiffer than they would ideally be to represent the NREL 5MW), and was put together by Frederik Berger at the University of Oldenburg. As we mentioned in Section 1, it is not usually possible to match the Reynolds number on a scaled wind turbine model—in the case of the MoWiTO, the blade chord Reynolds number is approximately 100 times lower than the Reynolds numbers seen on the NREL 5MW turbine. To account for this, the blades were redesigned to achieve a lift distribution similar to the NREL 5MW under the lower Reynolds numbers. For full details about the MoWiTO, see Berger et al. [12].

The MoWiTO operates in the test section of a large, closed-circuit wind tunnel at the ForWind – Center for Wind Energy Research at the University of Oldenburg [19]. The tunnel test section, opened officially in 2017, is $3 \times 3$ m$^2$ in cross-section and 30 m long in the open configuration used for our testing. The ForWind tunnel has an active grid, described by Kröger et al. [24] and based on previous developments [25, 26], over the inlet to the test section that allows for generating complex atmospheric conditions during testing. For example, the active grid can be used to produce gust wind conditions (used for testing in our first experimental campaign [13]) and turbulent inflow (used during our second campaign [14]), as well as step changes in wind speed, sheared boundary layers, and other inflows. Importantly, the active grid can reproduce the same 'random' flows, which is easy to achieve in simulation but impossible to achieve in the field. This allows for direct side-by-side comparisons of different controllers [13, 14] in complex wind conditions. To reduce the effect of shear from the floor, the inlet to the test section is raised 1.5 m above the ground; the MoWiTO is therefore mounted on a support structure to raise it to the height of the main flow. See Figure 1 for a schematic and photo of the main components of the testbed.

The real-time control for the MoWiTO was handled by a National Instruments CompactRIO-9066 (cRIO) running the LabVIEW controller (see Section 2.1). This cRIO has a 667 MHz dual-core CPU and runs the National Instruments Linux Real-time operating system. Our controller VI was embedded in a larger, high-level operating code developed at the University of Oldenburg, which, among other things, handles the start-up and shut-down procedures for the MoWiTO; data logging; and controller selection. This piece of code is extremely useful for rapidly testing different control laws on the MoWiTO essentially by simply replacing the controller VI within the higher-level operating VI.

During testing, the cRIO was placed on the support table behind the MoWiTO (white box in Figure 1), with electrical cabling run from the turbine nacelle to the cRIO via the interior of the tower. To test feedforward control approaches, we used a hot-wire measurement of the wind speed upstream of the turbine. The analogue hot-wire signal was amplified and transmitted to the cRIO via coaxial cable. The hot-wire anemometer needed to be calibrated at the beginning of each session and recalibrated periodically during longer testing sessions to ensure consistency in the measurements. Finally, the communications between the real-time cRIO and monitoring desktop in the control room were managed via ethernet.

We ran tests with a variety of both deterministic and turbulent wind sequences. For shorter sequences (gusts and short turbulent inflows), the inflow was repeated ten times for each controller. A timing signal was used to ensure that the turbine signals were correctly aligned between the controllers during data post-processing. For more details about the inflows used in our testing, see Sinner et al. [14]; for details about how the active grid is used to generate inflows, please refer to Kröger et al. [24] and Knebel et al. [26].

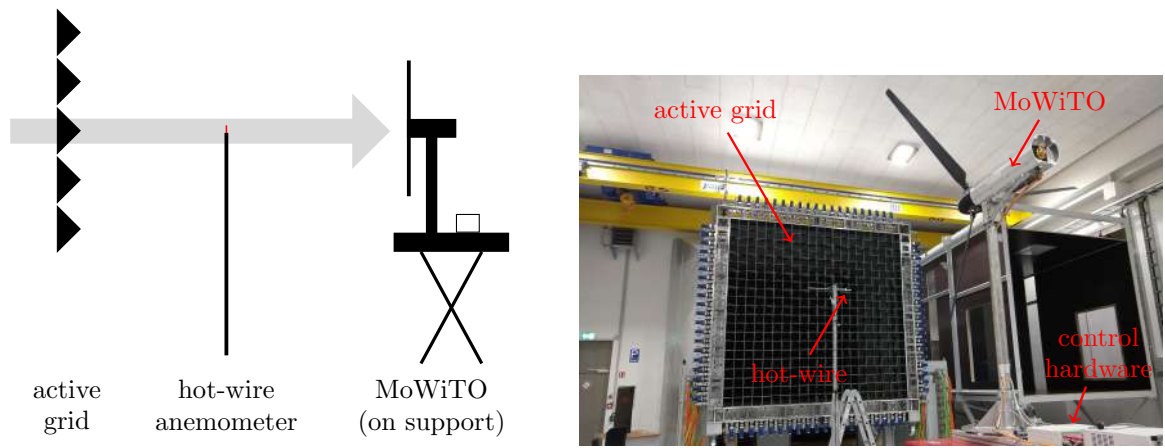

Figure 1: **Left:** Schematic of the MoWiTO wind turbine in the wind tunnel testbed. The gray arrow represents the wind flowing out of the inlet nozzle, through the active grid and across the MoWiTO. The white box on the support table represents the controller hardware that the controller was run on. **Right:** Labelled photograph of the experimental setup.

## 3 Conclusions and takeaways

Having documented our experimental approach, we will close by providing a handful of recommendations regarding different aspects of the approach. Some may seem obvious; others perhaps less so.

**Control model accuracy.** Determining how many of degrees of freedom to use in the mathematical model, and how carefully one needs to describe the modes included, is one of the major challenges faced by a control engineer and yet does not receive much attention in the literature. A higher order, more accurately identified model may lead to slight improvements in performance over a lower order, roughly identified model; however, one of the main benefits of feedback control is its robustness to plant uncertainty. Moreover, a simpler model is easier to work with and understand, and any improvement in performance from a more complex model may be negated due to a lack of understanding of the model behavior.

When testing advanced controllers on a physical testbed, we propose that the mathematical model be kept as simple as possible, at least initially. For our tests [13, 14], we modeled only the rotational mode of the turbine. This allowed us to 1) identify the model parameters very easily, 2) formulate the optimal control problem in a straightforward way, and 3) avoid the use of a state estimator. On the downside, we did notice that in some cases the main tower fore-aft mode was excited during testing, and for future work, a tower model could be worth including in the formulation.

We should also mention that ideally, software in the loop simulations and controller verification would take place with a plant model of higher fidelity than the mathematical control model. For our testing, we used a FAST model, which contains a nonlinear aeroelastic blade model on top of the main turbine rotational components.

**Controller simplicity.** Similarly, we suggest using a simple controller formulation. This is not to say that a simple control architecture should be used—the purpose of the testing we carried out is, after all, to test advanced controllers—but that choices should be made in the design of the chosen advanced controller that are easy to follow and understand and are reproducible. As with model selection, more complex tunings and choices may well lead to slight performance improvements; however the purpose of our testing, and we believe most of academic scaled-model testing, is to demonstrate that the controller works in a physical setting, rather than to maximize absolute performance.

Ideally, the complexity of the controller could be increased gradually. Our experiments took place over the course of two experimental campaigns, with the first investigating unconstrained linear-quadratic control [13] and the second investigating constrained linear-quadratic control [14]. Although this change meant that the controller implementation for the second experiments were considerably more complex than the first, mathematically the control problems were very similar; this meant that we could focus on the changes during the second experiment, rather than developing a model predictive controller from scratch. Even within the model predictive control framework, one could consider adding in constraints one at a time, perhaps beginning with simpler box input constraints and moving to affine input constraints and state constraints. We did not take that approach, but in hindsight, that may have helped us to understand where the main challenges lay in terms of computational speed.

**Controller implementation.** In this testing, we used a high-level language (Matlab/Simulink) for controller design and initial simulation testing before translating the design into a real-time code (LabVIEW/C++). While one could design the controllers directly in the real-time language, we find that the higher-level languages allow for quicker reworking of the code and therefore a quicker turnaround in design iterations. The downside to this is the extra step required to translate the code into the real-time language and then verify its behavior.

However the controller is implemented, we would like to underscore the value in hardware in the loop testing once the controller has been compiled on the real-time hardware. Having a suitable software testbed for HIL testing requires some effort; we were fortunate to have had the testbed set up beforehand, meaning that for us, HIL testing was very straightforward. If there are a series of physical tests planned, the upfront effort of developing a HIL testbed will likely be worthwhile.

**Preparing early.** As cliché as it is to say, being prepared early is perhaps the most important aspect of a test campaign. Often the window when the physical testbed is available is short, and it can be critical to keep that time open for addressing unforeseen issues in testing.

During our second campaign, the C++ code compilation for the real-time cRIO target was more complicated, and took considerably longer, than we had anticipated. For someone else, that step might have been straightforward, but more likely than not there will be challenges with the software that are best to overcome before the testing window opens. Once the testing window had begun, we had mechanical problems that needed to be addressed before the turbine was operational and responding correctly, limiting the amount of the testing window available for data collection. We were able to get everything done, but it would have been better if all of the software had been finalized and compiled before the testing window started, rather than having to work on both the software and hardware during that window. In the worst case, the testing simply finishes early!

Testing on scaled-model turbine testbeds is an important step in the process between theoretical control law design and industry acceptance, and we expect wind tunnel testing of turbine and farm controllers to continue to hold an important place in the literature. We hope that this documentation of our testing procedures and insights provides a useful reference for others that want to test controllers in physical, scaled-model testbeds, and we encourage readers to reach out to us with any questions that you have regarding our approach. We would be happy to provide further details where possible.

## Acknowledgements

The experiments described in this paper would not have been possible without the travel funding from the German Academic Exchange Service (DAAD) and the Hanse-Wissenschaftskolleg in Delmenhorst, Germany. The state of Lower Saxony supported the wind tunnel campaign, active grid, and MoWiTO under the project "ventus efficiens." We also thank the National Renewable Energy Laboratory and a Palmer Endowed Chair at the University of Colorado Boulder for ongoing financial support. Finally, we thank David Onnen for presenting this work in our absence.

## References

[1] Wiser R and Bolinger M 2021 Land-based wind market report: 2021 edition Tech. rep. U.S. Department of Energy, Lawrence Berkely National Laboratory

[2] Pao L and Johnson K 2011 *IEEE Control Systems Mag.* **31** 44–62

[3] Laks J H, Pao L Y and Wright A D 2009 Control of wind turbines: Past, present, and future *Proc. American Control Conf.* (St. Louis, MO) pp 2096–2103

[4] Scholbrock A, Fleming P, Schlipf D, Wright A, Johnson K and Wang N 2016 Lidar-enhanced wind turbine control: Past, present, and future *Proc. American Control Conf.* (Boston, MA) pp 1399–1406

[5] Lio W H, Rossiter J A and Jones B L 2014 A review on applications of model predictive control to wind turbines *UKACC Int. Conf. Control* (Loughborough, UK) pp 673–678

[6] Bossanyi E A, Fleming P A and Wright A D 2013 *IEEE Trans. Control Systems Technology* **21** 1067–1078

[7] Harris M, Hand M and Wright A 2005 Lidar for turbine control Tech. Rep. NREL/TP-500-39154 NREL Golden, CO

[8] Jonkman J, Butterfield S, Musial W and Scott G 2009 Definition of a 5-MW reference wind turbine for offshore system development Tech. Rep. NREL/TP-500-38060 NREL Golden, CO

[9] Jonkman J and Buhl Jr M 2005 FAST user's guide Tech. Rep. NREL/EL-500-38230 NREL Golden, CO

[10] Dickler S, Wintermeyer-Kallen T, Zierath J, Konrad T and Abel D 2021 *Forsch Ingenieurwes* **85** 313–323

[11] embotech 2016 Optimal wind turbine control online; accessed April 2021 URL https://www.embotech.com/wp-content/uploads/Vestas_story_embotech.pdf

[12] Berger F, Kröger L, Onnen D, Petrović V and Kühn M 2018 Scaled wind turbine setup in a turbulent wind tunnel *Deep Sea Offshore Wind R&D Conf.* (*J. Physics: Conf. Series* vol 1104) p 012026

[13] Sinner M, Petrović V, Berger F, Neuhaus L, Kühn M and Pao L Y 2020 Wind tunnel testing of an optimal feedback/feedforward control law for wind turbines *Proc. IFAC World Congress* (Berlin, Germany)

[14] Sinner M, Petrović V, Langidis A, Neuhaus L, Hölling M, Kühn M and Pao L Y 2021 *IEEE Trans. Control Systems Technology*

[15] Verwaal N, van der Veen G and van Wingerden J W 2015 *Wind Energy* **18** 385–398

[16] Campagnolo F, Weber R, Schreiber J and Bottasso C L 2020 *Wind Energy Science* **5** 1273–1295

[17] Navalkar S T, van Solingen E and van Wingerden J W 2015 *IEEE Trans. Control Systems Technology* **23** 2101–2116

[18] Fontanella A, Bayati I and Belloli M 2018 Control of floating offshore wind turbines: Reduced-order modeling and real-time implementation for wind tunnel tests *Proc. ASME Int. Conf. Offshore Mechanics and Arctic Engineering* (*Ocean Renewable Energy* vol 10) (Madrid, Spain)

[19] Petrović V, Berger F, Neuhaus L, Hölling M and Kühn M 2019 Wind tunnel setup for experimental validation of wind turbine control concepts under tailor-made reproducible wind conditions *Proc. Science of Making Torque from Wind* (*J. Physics: Conf. Series* vol 1222) p 012013

[20] Franklin G, Powell D and Emami-Naeini A 2019 *Feedback Control of Dynamic Systems* 8th ed (Pearson) ISBN 0134685717

[21] The Mathworks Inc MATLAB URL https://www.mathworks.com/products/matlab.html

[22] Ferreau H, Kirches C, Potschka A, Bock H and Diehl M 2014 *Mathematical Programming Computation* **6** 327–363

[23] National Instruments LabVIEW URL https://www.ni.com/en-us/shop/labview.html

[24] Kröger L, Frederik J, van Wingerden J W, Peinke J and Hölling M 2018 Generation of user defined turbulent inflow conditions by an active grid for validation experiments *Proc. Science of Making Torque from Wind* (*J. Physics: Conf. Series* vol 1037) p 052002

[25] Makita H 1991 *Fluid Dynamics Research* **8** 53–64

[26] Knebel P, Kittel A and Peinke J 2011 *Experiments in Fluids* **51** 471–481